\documentstyle[twoside, epsf]{article}
\begin{document}

\pagestyle{myheadings}
\markboth
{Toyoki Matsuyama and Hideko Nagahiro}
{Nara International Conference 1999}

\begin{center}
\LARGE Topological Mass vs. Dynamical Mass: \\
{\large {\bf Novelty in (2+1)-dimensional Space-Time}} 
\vskip 0.5cm
\large Toyoki Matsuyama 
\footnote{e-mail address: matsuyat@nara-edu.ac.jp} \\
Department of Physics, Nara University of Education \\
Takabatake-cho, Nara 630-8528, JAPAN \\
and \\
Hideko Nagahiro \\
Department of Physics, Nara Women's University  \\
Nara 630-8506, JAPAN
\end{center}

\begin{abstract}
The restriction of space-time dimensions to "2+1" leads us to a novel quantum 
field theory which has the Chern-Simons term in its action.  
This term changes the nature of gauge interaction by giving a so-called 
topological mass to a gauge field without breaking the gauge symmetry.  
We investigate how a dynamical mass of fermion is affected by the 
topological mass in the non-perturbative Schwinger-Dyson method.  
\end{abstract}

\section{Introduction}

Generally speaking, dimensions of space-time may become a critical restriction 
on an allowed law of interaction in that space-time.  
For example, the inverse-square law of gravitational and electromagnetic 
forces can be considered to be resulted from the fact that the dimensions of 
space-time is "3+1".  
This also is related to the fact that the Huygens' principle holds only 
in the (3+1)-dimensional space-time.~\cite{ken}  
The number "3+1", the dimensions of our space-time, is really a magic 
number.  

This kind of restriction may be expected in other space-time dimensions.  
The modern technology of engineering makes possible to produce low-dimensional 
electron systems in realistic electronic devises.  
Especially, in (2+1)-dimensional systems, novel phenomena as the Quantum 
Hall Effect~\cite{QH} and the High-$T_C$ Superconductivity~\cite{HTC} 
were discovered.  
It is plausible that these phenomena may have their origins in the dimensions 
of the space-time.  

In a sense, the dimensions "2+1" is more mysterious because the 
mathematics tells us that there exists a specific term called a 
Chern-Simons term~\cite{CS}.  
This term is allowed just in (2+1) dimensions.  
As is well known, the Lagrangian density for the electromagnetic field is 
given by the Maxwell term, which is (i) gauge invariant, (ii) Lorentz 
invariant, and (iii) bilinear for the gauge field.  
The Chern-Simons term also satisfies all of (i) $\sim$ (iii).  
Therefore the Maxwell theory has a natural extension which is defined by 
adding the Chern-Simons term to the Maxwell Lagrangian.  
This extended version is called the Maxwell-Chern-Simons 
theory.~\cite{SS,DJT}  
Thus the restriction of the space-time dimensions to "2+1" opens a pass to a 
new type of theory.  

In fact, there appeared many approaches to understand these macroscopic 
quantum effects by using (2+1)-dimensional quantum field theories with and 
without the Chern-Simons term.  
For example, the (2+1)-dimensional quantum electrodynamics ($QED_3$) has 
been used to explain the quantum Hall effect, and $QED_3$ with the 
Chern-Simons term (Maxwell-Chern-Simons $QED_3$) has provided the anyon 
model which is expected to give an essential mechanism for the high-$T_C$ 
superconductivity.~\cite{W}  
These investigations have produced important results and are now still 
in progress.  

What is the physical meaning of the Chern-Simons term?  
The Chern-Simons term gives the gauge field a mass without breaking the gauge 
symmetry.~\cite{SS,DJT}  
This mass is called a topological mass because the Chern-Simons term has a 
topological meaning as the secondary characteristic class.~\cite{CS}  
Whether the gauge field is massless or massive affects the nature of 
interactions, e.g., the range of interactions.  
In this case, the most important effect of the massive gauge field is to 
rescue the (2+1)-dimensional Maxwell theory from the infrared catastrophe 
which appears in a self-energy of fermion when the Maxwell field interacts 
with matters.~\cite{JTAPRS}  

On the other hand, it is known that nonperturbative radiative corrections can 
produce a mass of fermion called a dynamical mass.  
The dynamical mass generation of four-component fermions in $QED_3$ 
without the Chern-Simons term has been studied in Ref.\cite{PABCWABKW}.  
One four-component fermion is equivalent to two two-component fermions.
The mass term of the two-component fermion breaks the parity ($P$) symmetry 
while the one of the four-component fermion breaks $P \times 
Z_2({\rm flavour}$) symmetry.  
In Ref.\cite{HMH}, one of the present authors and others have investigated the 
dynamical mass generation of a single two-component fermion in $QED_3$ 
without the Chern-Simons term.  
A parity-breaking solution which generates the dynamical mass has been 
found .   

Both analyses have been extended to the cases with the Chern-Simons term.  
The study of the four-component dynamical mass in the Maxwell-Chern-Simons 
$QED_3$ has been done in Ref.\cite{HPKKKM}.  
The dynamical mass generation of a single two-component fermion in 
the Maxwell-Chern-Simons $QED_3$ has been studied in Ref.\cite{MNU}.  
Especially, Ref.\cite{MNU} is motivated to clarify a role of the topological 
mass in the nonperturbative dynamics.  

However, as was pointed out in Ref.\cite{MNU}, the estimation of the dynamical mass 
for a very small value of the topological mass is very difficult technically 
and also is highly nontrivial.  
In this paper, we extend further the analysis of Ref.\cite{MNU} to the case 
in which the topological mass has much more smaller value.  

This paper is organized as follows.  
In Sec. 2, we explain the Maxwell-Chern-Simons $QED_3$.  
A perturbative analysis is presented in Sec. 3.  
The Schwinger-Dyson equation is derived in Sec. 4.  
An approximated analytical analysis is done in Sec. 5.  
Results obtained by a numerical method is shown in Sec. 6.  
Finally we give conclusions with a discussion in Sec. 7.    

\section{Maxwell-Chern-Simons $QED_3$}

We consider the Maxwell-Chern-Simons $QED_3$ with the single two-component 
Dirac fermion. ~\cite{SS,DJT}  
The Lagrangian density of the theory is given by 
\begin{eqnarray}
{\cal L}= - \frac{1}{4} F_{\mu\nu} F^{\mu\nu}
          + \frac{\mu}{2} \varepsilon^{\mu \nu \rho} A_\mu \partial_\nu A_\rho
          - \frac{1}{2\alpha}(\partial_\mu A^\mu)^2
          + \bar{\psi}(i \not \! \partial - e \not \! \! A)\psi \ \ , 
\label{lag}
\end{eqnarray}
where $e$ is the gauge coupling constant and $\alpha$ is the gauge-fixing 
parameter.  
The second term in the right-hand side of Eq.(\ref{lag}) is 
the so-called Chern-Simons term.  
It is well-known that the term gives the gauge field the mass $\mu$ without 
breaking the gauge symmetry.  
In fact, a free propagator of the gauge field $iD_{\mu\nu}(p-k)$ derived from 
Eq.(\ref{lag}) is written as 
\begin{equation}
iD_{\mu\nu}(p)=
-i\frac{1}{p^2-\mu^2}\left(g_{\mu\nu}-\frac{p_\mu p_\nu}{p^2}\right)
+\mu\frac{1}{p^2-\mu^2}\frac{1}{p^2}\varepsilon_{\mu\nu\rho}p^{\rho}
-i\alpha\frac{p_\mu p_\nu}{p^4}.
\label{D}
\end{equation}
We find a massive pole at $p^2=\mu^2$.  
$\mu$ is called the topological mass.  

$\psi$ is the two-component fermion field which belongs to the irreducible 
spinor representation in (2+1)-dimensions.  
The Dirac matrices are defined by $\gamma^0=\sigma_3, \gamma^1=i\sigma_1, 
\gamma^2=i\sigma_2$ with diag$(g^{\mu\nu})=(1,-1,-1)$ where $\sigma_i$'s 
(i=1, 2, 3) are the Pauli matrices.  
The $\gamma^\mu$'s satisfy relations as 
$\{ \gamma^\mu, \gamma^\nu \}=2g^{\mu \nu}$, $\gamma^\mu \gamma^\nu = -i 
\epsilon^{\mu \nu \rho} \gamma_\rho + g^{\mu \nu}$ and $tr[\gamma^\mu 
\gamma^\nu ] = 2g^{\mu \nu}$.  
In this representation, there does not exist a matrix which anti-commutes 
with all of $\gamma^\mu$'s so that we cannot define the chiral 
transformation.  
This is a specific aspect of the odd-dimensional space-time.  
In even-dimensions, the chiral symmetry requires that a fermion is 
massless.  
In odd-dimensions, the chiral symmetry itself does not exist.  
Instead, the mass term of the fermion is forbidden by parity symmetry.  

Under the parity transformation
\footnote{In (2+1)-dimensions, the parity transformation is defined as \\
$ x=(t, x, y) \rightarrow x'=(t, -x, y) \ \ , 
\psi(x) \rightarrow \gamma^1 \psi(x') \ \ , 
A^0 (x) \rightarrow A^0 (x') \ \ , 
A^1 (x) \rightarrow - A^1 (x') \ \ , 
A^2 (x) \rightarrow A^2 (x') \ \ .  \nonumber
$
}, the mass term of the fermion and the Chern-Simons term change their signs.  
Thus the mass terms of both the fermion and the gauge field are forbidden by 
the parity symmetry.  
We study how the breaking of parity by the topological mass affects the 
mass generation of the fermion.  

\section{Perturbation}

Before proceeding to a nonperturbative analysis, it would be useful to see 
the lowest order of perturbation.  
The fermion self-energy in the one-loop approximation, $\Sigma^{(1)}(p)$, is 
expressed as 
\begin{equation}
\Sigma^{(1)}(p)=\int \frac{d^3k}{(2\pi)^3}
(-ie\gamma^\mu) iS_F(k)(-ie\gamma^\nu)
 iD_{\mu\nu}(p-k) \ \ ,
\label{self1}
\end{equation}
where $iS_F(p)$ is a free fermion propagator written as 
\begin{equation}
iS_F(p)=\frac{i}{\not{\hspace{-0.5mm}p}} \ \ ,
\label{freefermion}
\end{equation}
and $iD_{\mu\nu}(p-k)$ is a free propagator of the gauge field given 
in Eq. (\ref{D}).  

The allowed form of the fermion propagator in the relativistic theory is 
written as
\begin{equation}
iS^{(1)}_F(p)=\frac{i}{A^{(1)}(p)\not{\hspace{-0.5mm}p}-B^{(1)}(p)} 
             =\frac{i}{\not{\hspace{-0.5mm}p}-i\Sigma^{(1)}(p)} \ \ ,
\label{SF1}
\end{equation}
where $A^{(1)}(p)$ and $B^{(1)}(p)$ are functions of $\sqrt{p_\mu p^\mu}$, 
while $\Sigma^{(1)}(p)$ depends on $p_\mu$'s.  
$A^{(1)}(p)^{-1}$ is the wave function renormalization and 
$B^{(1)}(p)/A^{(1)}(p)$ is a mass induced by dynamical effects at the 
momentum scale $p$.  
The so-called dynamical mass $m_{phys}$ is defined by 
$m_{phys}=B^{(1)}(0)/A^{(1)}(0)$ as usual.
It is useful to notice the relations as 
\begin{equation}
tr\left[\Sigma^{(1)}(p)\right]=-2iB^{(1)}(p), \ \ 
tr\left[\not{\hspace{-0.5mm}p}\Sigma^{(1)}(p)\right]
=2i \{ A^{(1)}(p)-1 \} p^2.
\label{trace}
\end{equation}

We substitute Eqs.(\ref{freefermion}) and (\ref{D}) into 
Eq.(\ref{self1}) and use Eq.(\ref{trace}).  
Then we change the metric to the Euclidean one by the Wick rotation as 
$(k^0,\vec{k}) \rightarrow (ik^0,\vec{k})$ and $(p^0,\vec{p}) \rightarrow 
(ip^0,\vec{p})$.  
Then $k^2$ and $p^2$ are replaced by $-k^2=-(k^0)^2-(k^1)^2$ and 
$-p^2=-(p^0)^2-(p^1)^2$.  
After that,  we transform the integral variables $k^\mu$'s to the polar 
coordinates $(k, \theta,\phi)$.  
The angular integrations on $\theta$ and $\phi$ can be done explicitly.  
Finally we obtain 
\begin{eqnarray}
B^{(1)}(p)&=&\frac{e^2}{8\pi^2 p}\int^\infty_0 dk \frac{1}{k}
\left[\
- \frac{1}{\mu}(p^2-k^2)\ln\frac{(p+k)^2}{(p-k)^2}
\right. 
\nonumber \\
&+& \left. \frac{1}{\mu}(p^2-k^2+\mu^2)\ln\frac{(p+k)^2+\mu^2}{(p-k)^2+\mu^2}
\right] , 
\label{B1} \\
A^{(1)}(p)&=&1+\frac{e^2}{8\pi^2p^3}\int^\infty_0 dk \frac{1}{k}
\left[\ 
-2pk(\alpha+1) 
\right. 
\nonumber \\
&+& \left. \left\{\frac{1}{2\mu^2}(p^2-k^2)^2
+ \frac{1}{2}\alpha(p^2+k^2)\right\}\ln\frac{(p+k)^2}{(p-k)^2}
\right.
\nonumber \\
&+& \left.
\left\{\frac{1}{2}\mu^2-\frac{1}{2\mu^2}(p^2-k^2)^2\right\}
\ln\frac{(p+k)^2+\mu^2}{(p-k)^2+\mu^2}
\right] \ \ .
\label{A1}
\end{eqnarray}

The dynamical mass of fermion is defined in the infrared limit so that we 
are interested in the behaviour of $A^{(1)}(p)$ and $B^{(1)}(p)$ in this 
limit.  
In the region of $p \ll 1$, Eqs.(\ref{B1}) and (\ref{A1}) are written as
\begin{eqnarray}
B^{(1)}(p)&=&\frac{e^2}{\pi^2}\int^\infty_0 dk \left[\frac{\mu}{k^2+\mu^2}
+O(p^2) \right]\\
A^{(1)}(p)&=&1+\frac{e^2}{\pi^2}\int^\infty_0 dk
\left[\frac{1}{3}\left\{\frac{1}{k^2}\alpha-2\frac{\mu^2}{(k^2+\mu^2)^2}
\right\}
+O(p^2)\right].
\end{eqnarray}
This infrared approximation makes the integration on k possible and we have 
\begin{eqnarray}
B^{(1)}(0)=\frac{e^2}{2\pi}\frac{|\mu|}{\mu} \ , \ \ 
A^{(1)}(0)=1-\frac{e^2}{6\pi}\frac{|\mu|}{\mu^2}+\frac{e^2}{3\pi^2}
\frac{\alpha}{\epsilon} \ \ ,
\label{Per}
\end{eqnarray}
where $\epsilon$ is the infrared cutoff in the integration on $k$.  

It should be noticed that $B^{(1)}(0)$ depends on the sign of $\mu$.  
This also may be a specific aspect in (2+1)-dimensions.  
The dependence of $A^{(1)}(0)$ on $\mu$ shows that only the Landau gauge is 
free from the infrared divergence.  
On the other hand, $A^{(1)}(0)$ is singular at $\mu=0$ so that the theory 
with the Chern-Simons term may not be smoothly connected to the theory 
without the Chern-Simons term in the perturbation.  
This situation found in the perturbation motivates us to study the $\mu 
\rightarrow 0$ limit of the dynamical fermion mass by a nonperturbative 
method.  
This issue is extensively studied in the successive sections.  

\section{Schwinger-Dyson equation}

In this section, we proceed to a nonperturbative analysis, where we use 
the Schwinger-Dyson technique to evaluate the dynamical mass of the fermion.  
The Schwinger-Dyson equation for the fermion self-energy $\Sigma(p)$ is 
written as
\begin{eqnarray}
\Sigma(p)=(-i e)^2 \int \frac{d^3k}{(2\pi)^3} \ \gamma^\mu \ 
  i S'_F (k) \ \Gamma^\nu(k,p-k) \ i D'_{\mu\nu}(p-k) \ \ .
\label{SD}
\end{eqnarray}
$\Gamma^\nu(k,p-k)$ is a full vertex function and $D'_{\mu\nu}(p-k)$ is a 
full propagator of the gauge field.  
$S'_F$ is the full propagator of the fermion field which is  written as  
\begin{eqnarray}
i S'_F(p)=\frac{i}{A(p)\not \hspace{-0.8mm}p - B(p)}
        =\frac{i}{\not \hspace{-0.8mm}p-i\Sigma(p)} \ \ ,
\label{SF}
\end{eqnarray}
which includes the full correction beyond the perturbative fermion propagator 
given in Eq. (\ref{SF1}).  

To analyze Eq.(\ref{SD}) further, we need to introduce any suitable 
approximation.  
In this paper, we limit ourselves to use the lowest ladder approximation 
where the full propagator of the gauge field and the full vertex are replaced 
by the free propagator and the bare vertex respectively as 
\begin{eqnarray}
i D'_{\mu\nu}(p-k) \approx i D_{\mu\nu}(p-k) \ , \ \ 
\Gamma^\nu(k,p-k) \approx \gamma^\nu \ \ ,  
\label{ladder}
\end{eqnarray}
where $i D_{\mu \nu}$ has been given in Eq.(\ref{D}).  
Thus the Schwinger-Dyson equation in the lowest ladder approximation becomes 
\begin{eqnarray}
\Sigma(p)=(-i e)^2\int\frac{d^3k}{(2\pi)^3}
        \gamma^\mu \,i S'_F(k)\gamma^\nu \,i D_{\mu\nu}(p-k) \ \ .
\label{SDld}
\end{eqnarray}

We substitute Eqs.(\ref{D}) and (\ref{SF}) into Eq.(\ref{SDld}).  
Following the same steps as getting Eqs.(\ref{B1}) and (\ref{A1}), we finally 
obtain the coupled integral equations as 
\begin{eqnarray}
B(p) &=& \frac{e^2}{8\pi^2p} \int_{0}^{\infty}dk 
            \frac{k}{A(k)^2 k^2+B(k)^2} \left[ \left\{\alpha B(k) - 
            \frac{1}{\mu}(p^2-k^2) A(k) \right\} 
            \ln\frac{(p+k)^2}{(p-k)^2} \right. \nonumber \\
        &+& \left. \left\{\frac{1}{\mu}(p^2-k^2) A(k) +\mu A(k) 
         +2 B(k) \right\} \ln\frac{(p+k)^2+\mu^2}{(p-k)^2+\mu^2} \right] 
         \ \ , 
\label{B} \\
A(p) &=& 1+\frac{e^2}{8\pi^2p^3}\int_{0}^{\infty}dk
            \frac{k}{A(k)^2 k^2+ B(k)^2} 
            \left[ -2pk(\alpha+1) A(k) 
            \right. 
            \nonumber \\
        &+& \left. \left\{\frac{1}{2\mu^2}(p^2-k^2)^2 A(k) 
          + \frac{1}{\mu}(p^2-k^2) B(k) 
          + \frac{1}{2}\alpha(p^2+k^2) A(k) \right\} 
            \ln\frac{(p+k)^2}{(p-k)^2}  
            \right. 
            \nonumber \\
        &+& \left. \left\{\frac{1}{2}\mu^2 A(k) 
          - \frac{1}{2\mu^2}(p^2-k^2)^2 A(k)
          + \mu B(k) - \frac{1}{\mu}(p^2-k^2) B(k) \right\} 
            \right.
            \nonumber \\
   &\times& \left. \ln\frac{(p+k)^2+\mu^2}{(p-k)^2+\mu^2} \right]
            \ \ ,
\label{A}
\end{eqnarray}
which contain only the integration on the radial variable k.  
We solve these equation by an approximated analytical method and also 
numerically by using an iteration method in the following sections. 

\section{Approximated analytical method}

\subsection{$\mu\rightarrow$0 limit}

We can check easily that Eqs.(\ref{B}) and (\ref{A}) reduce to the 
Schwinger-Dyson equations in $QED_3$ without the Chern-Simons term if we 
put the topological mass $\mu$ equal to zero.  
In fact, taking the limit as $\mu \rightarrow 0$ in Eqs. (\ref{B}) and 
(\ref{A}), we obtain 
\begin{eqnarray}
B(p)&=&(\alpha+2) \frac{e^2}{8 \pi^2 p} \int^\infty_0 dk \frac{kB(k)} 
     {A(k)^2 k^2 + B(k)^2} \ln\frac{(p+k)^2}{(p-k)^2} \ \ , 
\label{BwoCS} \\
A(p)&=&1 - \alpha \frac{e^2}{4 \pi^2 p^3} \int^\infty_0 dk \frac{kA(k)} 
     {A(k)^2 k^2 + B(k)^2} \left[ pk - \frac{p^2+k^2}{4} 
      \ln\frac{(p+k)^2}{(p-k)^2} \right] ,
\label{AwoCS}
\end{eqnarray}
which are the Schwinger-Dyson equations in the lowest ladder approximation
derived in $QED_3$ without Chern-Simons term.  
We can see that there exists the specific gauge where the wave function 
renormalization is absent.  
Thus in the Landau gauge$(\alpha=0)$, Eq.(\ref{AwoCS}) gives us the simple 
solution as $A(p)=1$ and the problem reduces to solve Eq. (\ref{BwoCS}) 
with $A(p)=1$.  

In the case with the Chern-Simons term, as is seen in Eqs.(\ref{B}) and 
(\ref{A}), there does not exist such a specific gauge where the wave function 
is not renormalized.  
So far we cannot find a self-evident reason that the Landau is still specific 
in $QED_3$ with Chern-Simons term, it must be fair to study Eqs. (\ref{B}) 
and (\ref{A}) for various values of the gauge-fixing parameter $\alpha$.  

\subsection{Constant approximation} 

Before proceeding to a numerical analysis, it is very useful if 
we can estimate $A(0)$ and $B(0)$ analytically even under a fairly crude 
approximation.  
The kernels of these integral equations are dumped rapidly as the integral 
variable $k$ increases so that the contribution from $k \approx 0$ is the 
most dominant one in the integrals.  
We approximate $A(k)$ and $B(k)$ by $A(0)$ and $B(0)$ in the integrals.  
We call this approximation ''the $constant$ approximation''.  
Of course this approximation might be too crude for our purpose and we only 
use the result as reference in the numerical analysis.  
Under this approximation, we can perform the remaining radial integration 
and obtain
\begin{eqnarray}
B(0)=\frac{e^2}{2\pi} + \frac{e^2}{12\pi}\alpha \ , \ \ 
A(0)=1+\frac{2\alpha}{\alpha+6} \ \ , 
\label{BAcnst}
\end{eqnarray}
where we have considered the case of $\mu>0$.  

From Eq.(\ref{BAcnst}), we can see that the dependence 
of $B(0)$ and $A(0)$ on the gauge-fixing parameter, the coupling constant 
and the topological mass has the following peculiar features:
\begin{itemize}
\item[1)] Dependence on the gauge-fixing parameter \\
$B(0)$ depends linearly on $\alpha$.  
It is suggestive that $A(0)$ is singular at $\alpha=-6$ where B(0) vanishes.  
In the Landau gauge ($\alpha=0$), $A(0)=1$ and $B(0)=e^2/2\pi$.  
$A(0)=1$ is favourable for us because $A(p)=1$ means that the 
Ward-Takahashi identity is satisfied.  
\item[2)] Dependence on the coupling constant \\
$A(0)$ does not depend on $e^2$.  
It means that the deviation of $A(0)$ from 1 is independent of the coupling 
constant.  
This is crucially different from the perturbative result given by 
Eq.(\ref{Per}) where the deviation is proportional to $e^2$.  
On the other hand, $B(0)$ is proportional to $e^2$.  
\item[3)] Dependence on the topological mass \\
We recognize that  there is no dependence on the topological mass 
$\mu$ in Eq.(\ref{BAcnst}).  
In fact, if we apply the constant approximation to the case without the 
Chern-Simons term, we obtain the same results as Eq.(\ref{BAcnst}).
It means that the amount of the explicit parity breaking in the gauge 
sector by the topological mass does not affect the dynamical mass in 
the fermion sector in the constant approximation.  
\end{itemize}

Now we proceed to a more precise numerical evaluation in the next section.  

\section{Numerical method}

\subsection{Nontrivial solutions and gauge dependence}

We solve the two coupled integral equations (\ref{B}) and (\ref{A}) 
numerically by using a method of iteration.  
First we substitute trial functions into $A(k)$ and $B(k)$ in the right-hand 
sides of Eqs. (\ref{B}) and (\ref{A}) 
and then calculate the integrals numerically.  
The outputs so obtained, $A(p)$ and $B(p)$, are substituted back to the 
right-hand sides until the outputs coincide with the inputs.  
Finally we obtain convergent functions $A(p)$ and $B(p)$, which satisfy the 
integral equations, if there exist any solutions of Eqs. (\ref{B}) and 
(\ref{A}).

We have obtained the nontrivial solutions for the various values of the gauge 
parameter $\alpha$.  
It has been found that $A(p)$ is fairy close to 1 in the Landau 
gauge ($\alpha=0$).  
In the case of $QED_3$ without the Chern-Simons term, $A(p)$ is exactly equal 
to 1 in the Landau gauge under the lowest ladder approximation.  
However, in the case of $QED_3$ with the Chern-Simons term, there may be no 
apparent reason that $A(p)=1$ in the Landau gauge.  
It is surprising that the numerical calculation of so complicated integral 
equations results $A(p)\approx 1$ in the Landau gauge.  
There might be a simple reason for explaining a peculiarity of the Landau 
gauge.  

As a way of indicating to what extent gauge symmetry is broken by the 
bare vertex approximation, it is helpful to study the gauge invariant 
condensate $\langle\bar{\psi}\psi\rangle$ as a function 
of $\alpha$.  The condensate is defined by 
$\langle\bar{\psi}\psi\rangle =i \lim_{x\rightarrow 0} {\rm tr}S'_F(x)$ 
where $S'_F(x)$ is a propagator in real space-time 
coordinates.\footnote{By using the Fourier transformation and the Wick 
rotation, we obtain
\begin{eqnarray}
<\bar{\psi} \psi>= \frac{1}{\pi^2} \int^\infty_0 dk 
                   \frac{k^2 B(k)}{A(k)^2 k^2 + B(k)^2} \ \ .
\nonumber
\end{eqnarray}
}
We have found that the $\alpha$ dependence may be considered to be fairly 
weak.~\cite{MNU}  

Hereafter we present the results obtained in the Landau gauge.

\subsection{Dependence on the topological mass}

What we are most interested in is the dependence of the dynamical fermion 
mass on the topological mass of the gauge field. In the constant 
approximation, it has been shown that both $A(0)$ and $B(0)$ do not depend 
on the topological mass.  
Is this true in the more precise numerical evaluation?

We have studied the dependence of $A(0)$ on the dimensionless parameter 
$\hat{\mu}$ which is defined by $\hat{\mu}=\mu / e^2 $.  
We have found that the deviation of $A(0)$ from $1$ is less than 1 \%.  
We may say that $A(0)$ is almost equal to 1 in all region of $\hat{\mu}$.  
It means that the $\hat{\mu}$-dependence of $A(0)$ is extremely 
weak.~\cite{MNU}  

The $\hat{\mu}$-dependence of $B(0)$ is nontrivial.  
One of our motivations is to see how the Maxwell-Chern-Simons 
$QED_3$ is smoothly connected to $QED_3$ without the Chern-Simons term 
in the $\hat{\mu} \rightarrow 0$ limit.  
Our numerical calculation shows that very small meshes are needed 
to obtain reliable values of $B(0)$ in the region $\hat{\mu} \ll 1$.  
Because of the limitation of our machine ability, we take  another strategy 
in which $B(0)$ evaluated on meshes of zero width is estimated by an 
extrapolation from some $B(0)$'s on meshes of different finite widths.  
An example of the extrapolation is shown in Fig.1 where $B(0)$ on meshes of 
zero width is estimated by the curve fitting of three $B(0)$'s on meshes of 
finite widths.  
\begin{figure}[h]
\epsfysize=5cm
\centerline{\epsfbox{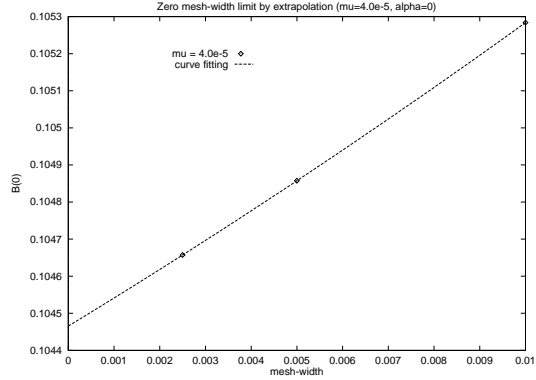}}
\caption
{The zero mesh-width limit of $B(0)$ by extrapolation.  
}
\end{figure}
\begin{figure}[h]
\epsfysize=5cm
\centerline{\epsfbox{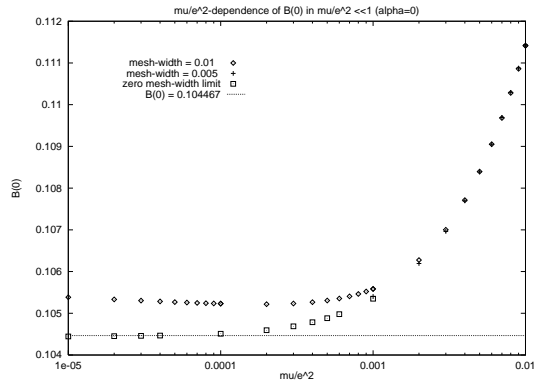}}
\caption
{The precise check of $\hat{\mu}$-dependence of $B(0)$ in the region of 
$10^{-5} \leq \hat{\mu} \leq 10^{-2}$.
}
\end{figure}

In Fig.2, we show the $\hat{\mu}$-dependence of $B(0)$ in the region 
$10^{-5} \le \hat{\mu} \le 10^{-2}$.  
$B(0)$'s on meshes of finite width show an abnormal behaviour in the region 
$\hat{\mu} \ll 1$ that $B(0)$'s depart from $B(0)$ of $QED_3$ without the 
Chern-Simons term as $\hat{\mu}$ decreases.  
This behaviour is improved by the extrapolation.  
$B(0)$'s obtained by the extrapolation smoothly tend to the value of $B(0)$ 
in $QED_3$ without the Chern-Simons term.  

The whole shape of $B(0)$ in the region $10^{-5} \leq \hat{\mu} \leq 10^4$ 
is given in Fig.3.  
$B(0)$ is almost constant in the region of $\hat{\mu} \gg 1$ and decreases 
rapidly in the region $\hat{\mu}= 1.0 \sim 0.01$.  
In the region of $\hat{\mu} \ll 1$,  $B(0)$ becomes almost constant again.  
The upper dotted line in Fig.3 is the result obtained in the constant 
approximation (Sec.5.2) and also in the lowest order perturbation (Sec.3).  
The lower dotted line shows the value obtained by a nonperturbative 
calculation in the case without Chern-Simons term.~\cite{HMH}  
\begin{figure}[h]
\epsfysize=5cm
\centerline{\epsfbox{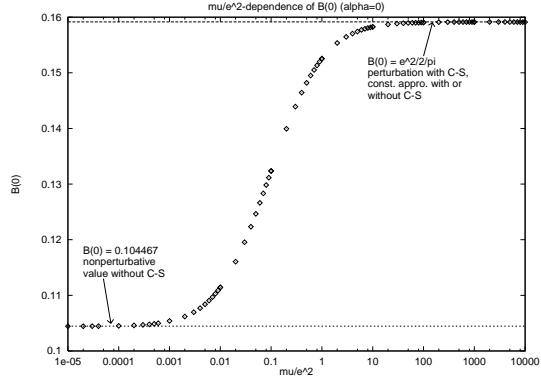}}
\caption
{$\hat{\mu}$-dependence of $B(0)$ in the region of $10^{-5} \leq \hat{\mu} 
\leq 10^4$.
}
\end{figure}

In other words, B(0) reproduces the result of perturbation in the region of 
$e^2 \ll \mu$, and in the region of $e^2 \gg \mu$, B(0) is close to the 
nonperturbative value obtained in the case without Chern-Simons term.

\section{Conclusion and discussion}

We have studied the dependence of the dynamical fermion mass on the 
topological mass in the Maxwell-Chern-Simons $QED_3$ nonperturbatively by 
using the Schwinger-Dyson method.  
When the topological mass is larger than the square of the coupling constant, 
the value of the topological mass remains to be the one obtained by the 
perturbation.  
As the topological mass decreases, the value is changed to a nonperturbative 
value rapidly.  
The transition from the perturbative value to the nonperturbative one is sharp 
but not critical.  
Though it seems not to be a phase transition, the inclusion of the 
Chern-Simons term changes the nature of the theory drastically.  

The most plausible application of our results can be found in condensed matter 
physics.  
It will be considered extensively in forth coming papers.  
What we would like to stress in this opportunity is an application to a 
model in the early Universe.  
In a study of space-time evolution of the Universe, the finite temperature 
effect may become important.  
In an imaginary time formalism, an inverse of the time dimension with a 
periodic (anti-periodic) boundary condition for a boson (fermion) is 
interpreted as a temperature.  
Then in a high-temperature limit, the time dimension vanishes and a 
(3+1)-dimensional space-time collapses into 3-dimensional space in which 
the Euclidean version of (2+1)-dimensional models may become effective.  
It also would be interesting to consider a mechanism which can produce the 
Chern-Simons term in the high temperature limit.  

One of the authors (T. M.) would like to thank Professor M. Kenmoku for his 
hospitality at the Nara Women's University.

\end{document}